\documentclass[runningheads]{llncs}
\pdfoutput=1
\usepackage[utf8]{inputenc} 
\usepackage[T1]{fontenc}    
\usepackage{hyperref}       
\usepackage{url}            
\usepackage{booktabs}       
\usepackage{amsfonts}       
\usepackage{nicefrac}       
\usepackage{microtype}      
\usepackage{xcolor}         
\usepackage{textcomp}

\usepackage{booktabs} 
\usepackage{amsmath,amssymb} 
\usepackage{graphicx}
\usepackage{multirow}
\usepackage{soul,color}
\usepackage{bm}
\usepackage{dsfont}
\usepackage{enumitem}
\usepackage{algorithmic}
\usepackage[linesnumbered,ruled,vlined]{algorithm2e}
\usepackage{booktabs}
\usepackage{caption}
\usepackage{url}
\usepackage{cite}
\usepackage{wrapfig, blindtext}
\usepackage{etoolbox}
\usepackage{bbm}
\usepackage{float} 
\usepackage{makecell}
\usepackage{setspace}
\usepackage{amsfonts}
\usepackage{subfig}

\begin{document}
\title{Low-Parameter Federated Learning with Large Language Models}
\titlerunning{Low-Parameter Federated Learning with Large Language Models}
%

\author{Jingang Jiang  \and Xiangyang Liu \and
Chenyou Fan}
\authorrunning{J. Jiang et al.}
%

\institute{South China Normal University, Guangdong, China \\
\email{\{2022024923,2022024952\}@m.scnu.edu.cn, fanchenyou@scnu.edu.cn}}

\maketitle              
\begin{abstract}
We study few-shot Natural Language Understanding (NLU) tasks with Large Language Models (LLMs) in federated learning (FL) scenarios. It is a challenging task due to limited labeled data and communication capacities in FL, especially with mobile devices. Recent studies show LLMs can be prompted to perform few-shot NLU tasks like sentiment analysis and arithmetic reasoning. However, the huge sizes of LLMs result in high computation and communication costs, making classical FL schemes impractical. To address these challenges, we propose Low-Parameter Federated Learning (LP-FL). LP-FL combines few-shot prompt learning from LLMs with efficient communication and federating techniques. Our approach enables federated clients to assign soft labels to unlabeled data using gradually learned knowledge from the global model. Through iterative soft-label assigning, we continually expand the labeled set during the FL process. Additionally, to reduce computation and communication costs, LP-FL utilizes the Low-Rank Adaptation (LoRA) technique for compact learnable parameter construction, efficient local model fine-tuning, and affordable global model federation. LP-FL consistently outperforms Full-Parameter Federated Learning (FP-FL) in sentiment analysis tasks across various FL settings. Its resistance to over-fitting allows LP-FL to equal or surpass centralized training in few-shot scenarios.


\end{abstract}

\section{Introduction}

The advent of Large Language Models (LLMs) such as GPT-3~\cite{brown2020language} has a profound impact on the research landscape of not only Natural Language Processing (NLP) studies but also the entire AI and Big Data communities. The fine-tuning of Large Language Models (LLMs) has proven highly effective across a multitude of tasks. In the current computing environment, the proliferation of mobile devices and sensors has become instrumental in information acquisition. Leveraging LLMs to process this wealth of information holds the potential to significantly enhance the convenience and expediency of our daily lives. However, the immense scale of parameters in LLMs entails significant computational costs for fine-tuning. Performing fine-tuning directly on these devices is impractical due to the formidable requirement of computation power. Hence, the pursuit of an effective fine-tuning method that achieves desirable outcomes with minimal parameter fine-tuning has become imperative.

Additionally, not all mobile devices can collect a sufficient amount of data. Typically, mobile device users only have processing permissions for the data generated during their own usage, and this data is often unlabeled. For example, mobile device users provide reviews for movies, hotels, or restaurants, and we can obtain a small amount of labeled data based on their positive or negative feedback. However, there is still a significant amount of comment data, such as those on TikTok or Twitter, where we cannot ascertain the sentiment. Fortunately, there is a substantial user base of mobile devices, which means that we have a significant amount of data available, albeit dispersed among various devices. Hence, distributed learning with few-shot labeled data has become an emerging topic. 

Federated learning (FL) enables distributed training of a global model on decentralized data. Hence, we propose leveraging FL to collaboratively train these devices and achieve effective fine-tuning of a global model. Additionally, we explore optimal utilization of labeled and unlabeled samples under the FL scenarios. One approach, PET~\cite{schick2020exploiting}, rephrases input examples using diverse prompts to aid the LLMs' understanding of the task. Fine-tuning is performed on each prompt using a LLM. The resulting models assign soft labels to unlabeled data, expanding the labeled dataset for standard supervised training. However, PET's multi-task setup in a federated environment incurs high computational costs and communication overhead, posing challenges for resource-constrained clients like mobile devices and sensors. 

Our contributions are summarized as follows:
\begin{enumerate}[noitemsep,leftmargin=*]

\item We consider an under-studied task of fine-tuning LLMs with distributed devices with limited communications and local computational powers. 

\item  We fine-tune the LLMs by adding task descriptions to the input examples for text sentiment classification. We start with a small number of labeled samples and then use a semi-supervised method to augment the dataset and enhance the fine-tuning process.

\item We introduce a low-parameter methodology called Low Parameter Federated Learning, abbreviated as LP-FL, for efficiently fine-tuning a small subset of the local model parameters then federate averaging over all clients.

\item We demonstrate that our method achieves comparable or even better performance than Full-Parameter Federated Learning, while greatly reducing computational costs and communication requirements on individual devices.
\end{enumerate}
\section{Related Work}
\textbf{Federated learning(FL). }~\cite{fedavg, zhao2018federated, sattler2019robust, li2019federated, wu2020personalized} is a distributed learning method that aims to train a global model on decentralized data while preserving data privacy. In FL, clients download copies of the global model and compute local gradients using their private data in each round. A central server coordinates the distributed clients, aggregating the local parameters to update the global model without exchanging raw data. Our research focuses on federated few-shot learning, addressing the challenge of training an effective global model in a federated environment with limited client-side labeled data. 

Several approaches have been proposed to tackle this scenario. FedFSL~\cite{fan2021federated} performs classification of unseen data classes using only a small number of labeled samples. FedSSL~\cite{fan2022private} utilizes semi-supervised learning to fully leverage labeled and unlabeled data sources for training. pFedFSL~\cite{zhao2022personalized} identifies well-performing models on specific clients (without revealing local data to the server or other clients) and selects suitable clients for collaboration, enabling personalized and distinctive feature space learning for each client. FedAffect~\cite{shome2021fedaffect} updates the feature extractor network on disjoint unlabeled private facial data to learn robust and diverse facial representations. Similarly, our work focuses on utilizing the power of pre-trained models to address this issue. Another approach, AUG-FedPrompt~\cite{cai2022aug}, performs data augmentation by annotating a large amount of unlabeled data, achieving operations similar to full fine-tuning with minimal initial labeled data. However, this method entails high communication costs due to full-parameter fine-tuning of the model.

\textbf{Large Language Models (LLMs). }
Pre-trained language models (PLMs) based on the transformer architecture, such as BERT~\cite{devlin2018bert} and GPT~\cite{radford2018improving}, have significantly enhanced the performance of natural language processing tasks. Expanding on this foundation, researchers continue to explore the upper limits of language model parameter sizes to unlock the potential of PLMs. The emergence of Large Language Models (LLMs) like GPT-3~\cite{brown2020language}, PaLM~\cite{chowdhery2022palm}, LaMDA~\cite{thoppilan2022lamda} and LLaMa~\cite{touvron2023llama} exemplifies this trend. These models exhibit remarkable few-shot capabilities by leveraging natural language prompts and task demonstrations as contextual input. However, these capabilities are built upon LLMs with parameter counts often exceeding 10 billion, making their application in real-world scenarios challenging. Consequently, researchers have begun investigating prompt performance on smaller-scale language models. Relevant approaches include PET~\cite{schick2020exploiting}, which reformulates input examples into cloze-style phrases to facilitate the language model's comprehension of the given task. AutoPrompt~\cite{shin2020autoprompt} selects a subset of discrete characters as triggers through gradient-based search and constructs templates to predict the probability of corresponding label words using models. LM-BFF~\cite{gao2020making} introduces a generative method for pattern construction, followed by a search-based technique to derive the associated verbalizers. Additionally, PPT~\cite{gu2021ppt} employ continuous prompts and transform templates into continuous vectors for optimization.

\textbf{Parameter-Efficient Fine-Tuning(PEFT). } PEFT enables efficient adaptation of LLMs to various downstream tasks without the need to fine-tune all parameters of the LLMs. The PEFT method fine-tunes only a small subset or additional parameters, while keeping the majority of pre-training parameters fixed, resulting in significant reductions in computation and storage costs. Notably, advanced PEFT techniques achieve performance comparable to full fine-tuning. Prefix-Tuning~\cite{li2021prefix} introduces a continuous and task-specific sequence of vectors, known as a prefix, added before the model input. This approach fixes all parameters of the LLMs and focuses on updating and optimizing the prefix specific to the task, resulting in minimal additional computational and storage overhead in downstream tasks. P-Tuning~\cite{liu2021gpt} leverages a limited number of continuous embedding parameters as prompts to optimize the performance of GPT in natural language understanding (NLU) tasks. In contrast, Prefix-Tuning is tailored for natural language generation (NLG) tasks. Moreover, P-Tuning solely introduces parameters in the embedding layer, whereas Prefix-Tuning incorporates trainable parameters across all layers. In contrast, P-Tuning V2~\cite{liu2021p} optimizes and adapts to NLU tasks by employing continuous prompts and updating prompt parameters at each layer of the model.Prompt Tuning~\cite{lester2021power} fixes all parameters of the LLMs and allows for the addition of a small, task-specific set of $k$ tokens to be prepended to the input text for each downstream task. As the model size reaches a certain scale, Prompt Tuning alone proves sufficient to achieve fine-tuning performance. Adapter Tuning~\cite{houlsby2019parameter, he2021towards, lin2020exploring, pfeiffer2020adapterfusion, ruckle2020adapterdrop} involves introducing new network layers or modules within the internal network layers of the LLMs to adapt to downstream tasks. Low-Rank Adaptation (LoRA)~\cite{hu2021lora}, while freezing the original model parameters, incorporates additional network layers and exclusively trains the parameters of these newly added layers to achieve results similar to full-model fine-tuning.

\begin{figure}[ht]
    \centering
    \includegraphics[width=1\textwidth]{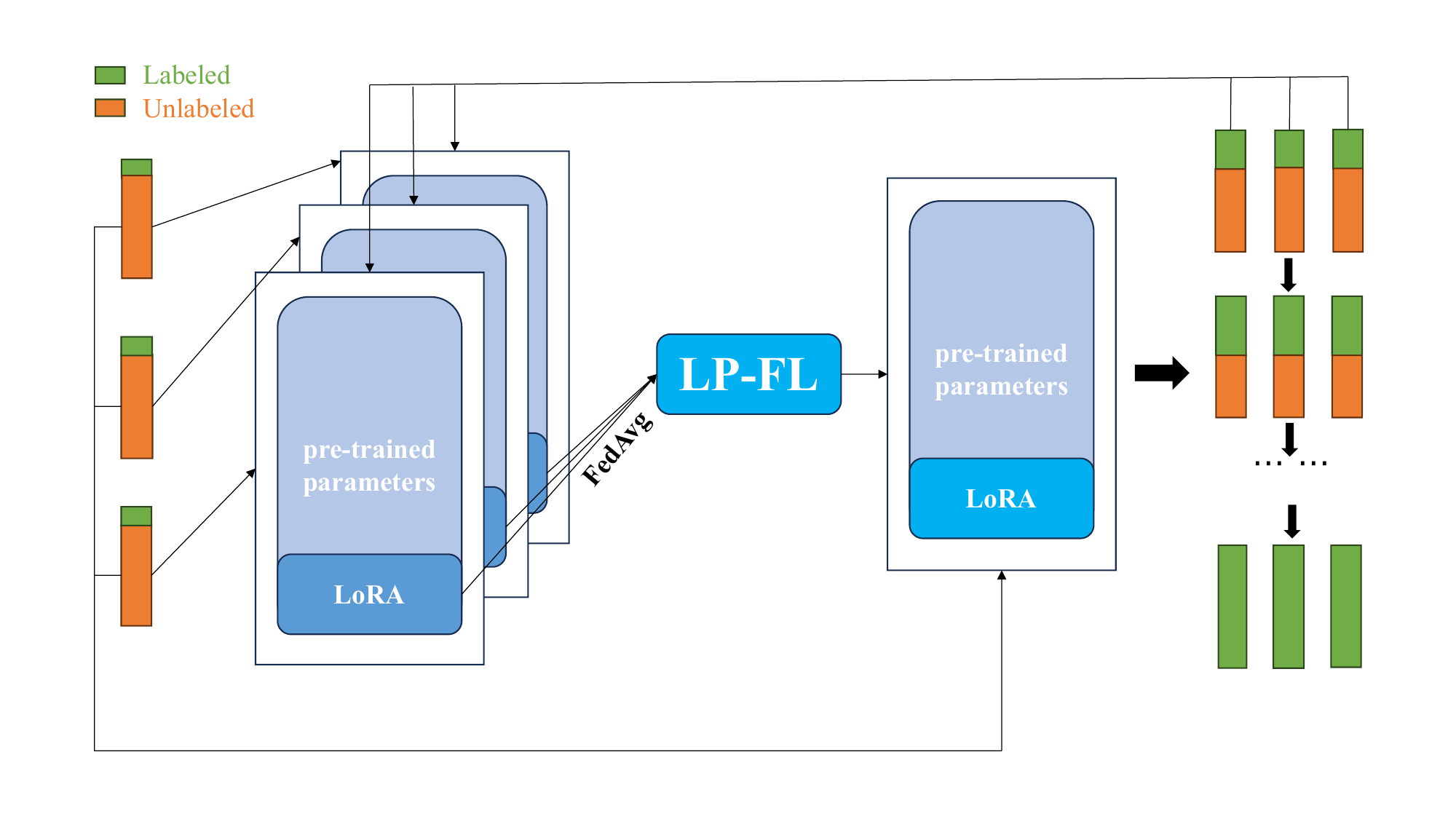}
    \caption{the training workflow of LP-FL:(1) Each client conducts fine-tuning of the global model using their respective local labeled data. The global model is a LLM with a LoRA bottleneck module, where only the parameters of LoRA can be trained. (2) After training, each client transfers their trained LoRA parameters to the server. The server performs FedAvg on the LoRA parameters and subsequently retransmits the updated LoRA parameters to each client. (3) Each client updates their local model using the received LoRA parameters and selects a portion of data from their local unlabeled dataset for annotation, thereby expanding the labeled dataset for further training continuation.}
    \label{fig:introduction}
\end{figure}

\begin{figure}[ht]
    \centering
    \includegraphics[width=1\textwidth]{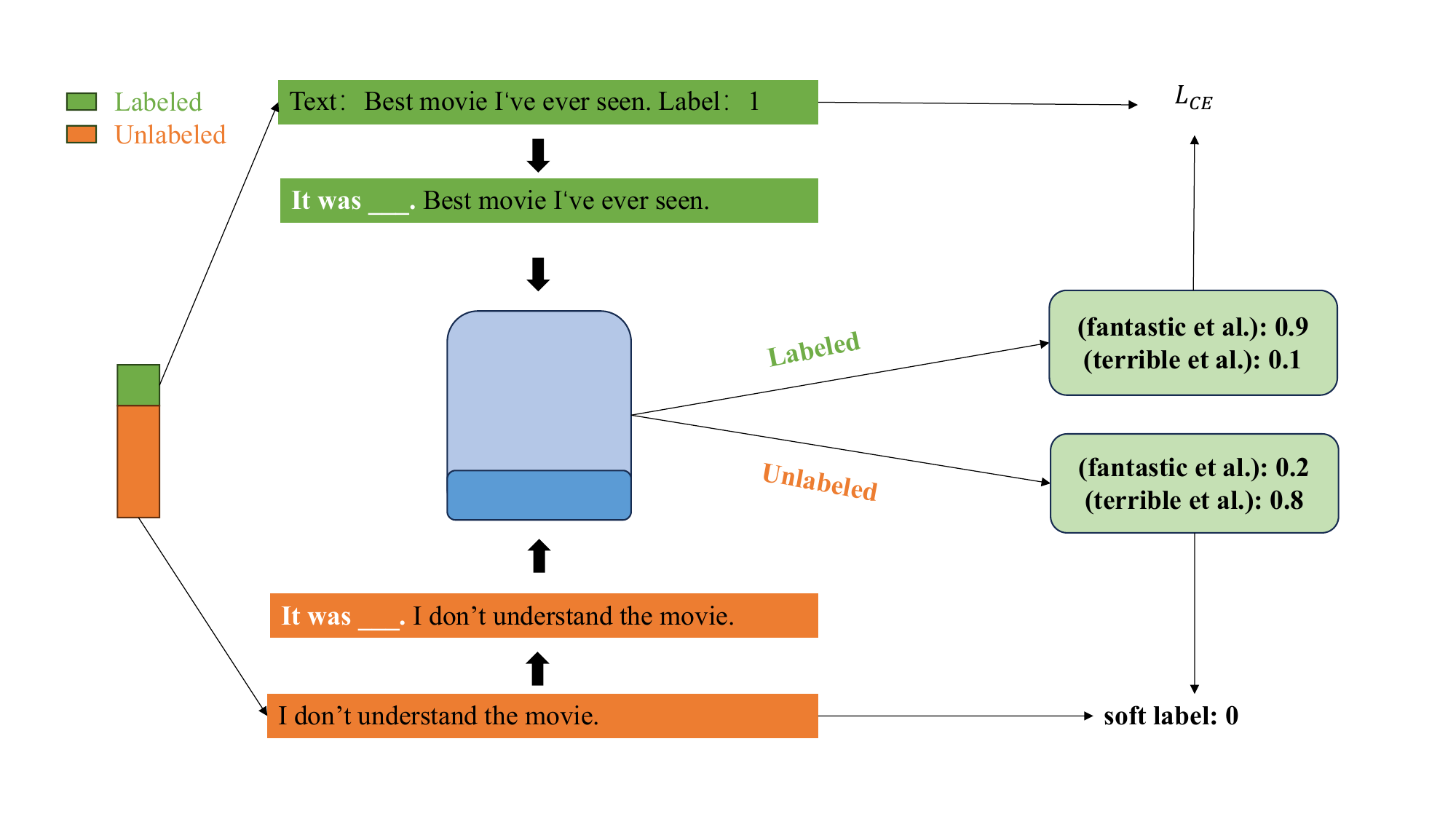}
    \caption{LP-FL Fine-tuning and Data Annotation Process: For each data point in the local dataset, we process it using a task description resembling "It was [MASK]." with a mask token. Subsequently, we utilize the LLM to predict the logits of all words in the label mapping at the masked position and convert them into classification logits. In the case of labeled data, these logits can be employed to compute the cross-entropy loss for fine-tuning the LLM. Conversely, for unlabeled data, soft labels can be generated to expand the labeled dataset.}
    \label{fig:task_description}
\end{figure}

\section{Approaches}
We describe the training workflow of Low-Parameter Federated Learning (LP-FL) in Figure~\ref{fig:introduction}. Assuming $M$ as a Large language model (LLM) with a vocabulary size of $V$ and the mask token [MASK] $ \in V$. Now we consider a federated environment with $K$ participating clients. As shown in Figure~\ref{fig:task_description}, each client $k$ possesses a labeled dataset $T_k$ comprising $n_k$ data instances, along with a much larger unlabeled dataset $U_k$ with $u_k$ instances, where $n_k \ll u_k$. For each client, given an input sequence example $x=(s_1,...,s_n)$ where each word $s_j \in V$, we can employ the $P(x)$ to add a task description phrase with a mask token to the input sequence, allowing the LLM to predict the word for the mask position. Using the mapping $L \rightarrow V$, we associate the label $l \in L$ of $x$ with a word $v \in V$. 

According to the above definition, our goal is to leverage the LLM $M$ to predict the probability of each $v \in V$ at the position of the mask token in $P(x)$. For an input sequence example $x$, we compute the score of $x$ with label $l \in L$ as follows:
\begin{equation}
s_P(l|x)=\sum_{v \in V}M(v|P(x)) \ .
\end{equation}
Upon acquiring the probability distribution over the labels through a softmax function, we can compare the predicted probability of the true word and measure the prediction with a standard cross-entropy loss, given below:
\begin{equation}
L_{CE} = \frac{1}{n} \sum_{i = 1}^n(-\log \frac{e^{s_P(l|x)}}{\sum_{l' \in L}e^{s_P(l'|x)}}) \ .
\label{loss}
\end{equation}
Give CE loss in Eq.(\ref{loss}), we can fine-tune the LLM $M$. 

For each client, we utilize multiple task descriptions $P \in \mathcal{P}$ to fine-tune the local model with the aforementioned approach. Furthermore, to address the communication burden arising from the substantial parameter size of the LLM, we utilize the Parameter-Efficient Fine-Tuning technique known as Low-Rank Adaptation (LoRA)~\cite{hu2021lora} during the fine-tuning phase. 

LoRA involves preserving the parameters of the LLM itself without modification. Specifically, LoRA introduces a fine-tuning-only bottleneck module which forms a residual connection to its original parameters $W_0$ to the original LLM. This bottleneck module is composed of two matrices A and B. Matrix $A$ reduces the input dimension from $d$ to $r$, then matrix $B$ restores the output dimension from $r$ to $k$, thereby simulating the concept of intrinsic rank, as shown in Eq.(\ref{lora}). 
\begin{equation}
\begin{split}
&W_0 + \Delta{W} = W_0 + BA, \\
s.t. \quad &  B \in \mathcal{R}^{d \times r}, A \in  \mathcal{R}^{r \times k}, r \ll min(d,k) \ . 
\end{split}
\label{lora}
\end{equation}

Throughout the training period, the pre-trained $W_0$ remains fixed and does not undergo gradient updates. The trainable parameters are the bottleneck parameter $A$ and $B$. In our work, we fixed the $r$ as $8$, which the trainable parameters will be about 30\% of the all parameters. Following the completion of a global training round, each client exclusively uploads $\Delta W$ to the server for parameter averaging. We adopt FedAvg as the method for parameter averaging as follows:
\begin{equation}
\min_{\Delta{W}\in \mathbb{R}^{d \times k}} \sum_{k=1}^{K}\frac{n_k}{n}F_k(\Delta{W}), \quad s.t. \quad F_k(\Delta{W})=\frac{1}{n_k}\sum_{i\in P_k}f_i(\Delta{W}) \ ,
\end{equation}
in which $f_i(w)$ is the loss function for client $i$, and $n$ is the total number of clients in the system. 

We employ an iterative approach wherein models are trained on continuously expanding datasets over multiple epochs. Each client assigns soft labels to a subset of their local unlabeled data using the global model after FedAvg. When labeling the unlabeled data $x \in U$, the prediction accuracy $a_P$ of each task description $P$ on the validation set is utilized as a weight. The soft label for $x$ is then obtained by performing a weighted average, which as follows:
\begin{equation}
s(l|x) = \frac{1}{Z} \sum_{P \in \mathcal{P}} a_P \cdot s_P(l|x), \quad s.t. \quad Z = \sum_{P \in \mathcal{P}} a_P \ .
\end{equation}

The model is subsequently fine-tuned on both the original labeled data and the newly annotated data, iterating through all rounds of training. 
We provide the algorithm details in~\ref{algo:Fed-LoRA} to illustrate the training procedures.

{
\begin{algorithm}[htp]
\small
\DontPrintSemicolon
\caption{\textbf{LP-FL overview}. ${M}$ is a global LLM; $T$ is the local labeled data, $U$ is the local unlabeled data ($|T| \ll |U|$); $l$ is step size, $G$ is the number of global rounds; $E$ is the number of  epochs.} \label{algo:Fed-LoRA}
\textbf{Server executes:} \;
\begin{Indp} Initialize global LLM ${M}$ with LoRA bottleneck modules, and only the parameters of LoRA bottleneck modules $w$ can be trained;

     $l \leftarrow 5 \times 10^{-5}$; \ $G \leftarrow 5$; \ $E \leftarrow 5$; \

\While{g $\leq$ $G$}{

    \For{\textup{each client} $k$ in $K$ clients \textup{\textbf{in parallel}}}{ 
    $w_{g+1}^k \leftarrow $ ClientUpdate $(k, w_g)$ \;
    }
    $w_{g+1} \leftarrow FedAvg(w_{g+1}^{1:K}) $  \;
    
    Server sends $w_{t+1}$ back to clients \;
    $g \leftarrow g+1$ \;
}
Return $M$ \;
\;

\end{Indp}
\textbf{ClientUpdate}$(k, w)$: // \textit{Run on client k} \;
\begin{Indp}
    $T$ is the local labeled data, $U$ is the local unlabeled data
    
    \eIf{$g < 5$} {
    $S \leftarrow $ label the $25\%$ of the $U$ \;
    $T \leftarrow T + S$ \;
    $U \leftarrow U - S$
    }{
    continue
    }
    
    $\mathcal{B} \leftarrow $ (split T and U into batches of size $B$) \;
    
    \For{\textup{each local epoch} $i$ \textup{from 1 to} $E$}{
        \For{\textup{batch b} $\in \mathcal{B}$ }{
            $w \leftarrow w$ - $\eta \nabla \ell(w;b)$
        }
    }

Return $w$ \;
\end{Indp}
\end{algorithm}
}

\section{Experiment}
\label{sec:exp}
We verify our approach on the widely used IMDB and Yelp datasets for evaluating sentiment analysis task. We performed an analysis of our method in both federated and centralized training settings, as well as an analysis of Low-Parameter Federated Learning (LP-FL) and Full-Parameter Federated Learning (FP-FL) in the federated environment. 
\subsection{Datasets and Task Description}
\label{sec:dataset}

\textbf{IMDB} dataset is a benchmark dataset for sentiment classification task. It contains tens of thousands of textual reviews specifically related to movie evaluations. We employed the complete training set, which contained 25,000 labeled data points. Nonetheless, during the experiments, only a portion of the data, not surpassing 10\%, was utilized as the labeled dataset, while the remaining data was designated as the unlabeled dataset. IMDB provides a well-balanced collection of movie reviews for sentiment classification tasks, which is widely compared in recent studies. We utilize the following task descriptions to process the text $x$ from the IMDB dataset:
\begin{align*}
P_1(x) &= \text{It was [MASK]. x.} \\
P_2(x) &= \text{Just [MASK]! x.} \\
P_3(x) &= \text{x. All in all, it was [MASK].} \\
P_4(x) &= \text{x. In summary, the movie is [MASK].}
\end{align*}
We conducted label mapping for the dataset as well:
\begin{align*}
v(0) = \text{[} &\text{'boring', 'disappointing', 'terrible', 'predictable', 'obvious', } \\
&\text{'dull', 'commonplace', 'awful', 'simple', 'confusing'}\text{]} \\
v(1) = \text{[} &\text{'good', 'great', 'excellent', 'funny', 'interesting', } \\
&\text{'amazing', 'fantastic', 'awesome', 'nice', 'inspiring'}\text{]}
\end{align*}

\textbf{Yelp} dataset is widely used for sentiment classification task~\cite{schick2020exploiting, gao2020making} collected from the Yelp platform. The dataset contains business details such as names, locations, categories, and star ratings. It also consists of user profiles with IDs, names, and review histories. The reviews themselves include text content, ratings, and additional metadata. Our study focuses on the yelp-polarity dataset, which is a binary sentiment classification dataset. Likewise, we conducted a random sampling of 25,000 labeled data points from the dataset. During the experiments, only a small subset of this data was employed as the labeled dataset, leaving the remaining samples designated as the unlabeled dataset. Below are the task descriptions for text $x$ in yelp-polarity dataset:
\begin{align*}
P_1(x) &= \text{It was [MASK]. x.} \\
P_2(x) &= \text{Just [MASK]! x.} \\
P_3(x) &= \text{x. All in all, it was [MASK].} \\
P_4(x) &= \text{x. In summary, the restaurant is [MASK].} 
\end{align*}
And the label mapping:
\begin{align*}
v(0) = \text{[} &\text{'dirty', 'bad', 'terrible', 'rude', 'obvious', } \\
&\text{'dull', 'commonplace', 'awful', 'simple', 'bland'}\text{]} \\
v(1) = \text{[} &\text{'good', 'cozy', 'inviting', 'delicious', 'impressive', } \\
&\text{'clean', 'organized', 'awesome', 'nice', 'fabulous'}\text{]}
\end{align*}

Our method is validated on the IMDB and Yelp datasets. The complete sets of 25,000 training and testing samples are used for IMDB, while we randomly select 25,000 samples from the training and testing sets for Yelp. The Large Language Model (LLM) employed in our experiments is BERT-Large model with 336M parameters. 

Considering previous research and practical considerations, we choose the following hyperparameter values: a batch size of 8, local epochs as 5, global training rounds as 5, a learning rate of $5 \times 10^{-5}$, a maximum sequence length of 128, and a rank of 8 for Low-Rank Adaptation (LoRA). We conduct experiments in federated settings with 2, 5, and 10 participating parties. During the training process, we selectively employ a fraction of the training dataset, specifically 1\%, 5\%, and 10\%, as labeled datasets. Consequently, each federated participant receives an equal distribution of total 250/1250/2500 labeled samples. Meanwhile, to establish the validation set for the global model, we randomly sample 1000 instances from the remaining data. The left 23750/22750/21500 samples are further allocated evenly to each  participant as the unlabeled data. 

Following each round, once all participants have concluded their local training, the LoRA parameters are transmitted to the server. The server employs FedAvg to update the LoRA parameters in the global model and redistributes the updated parameters to all participants for local model updating. During this phase, participants employ the updated model to randomly annotate 25\% of the data from the unlabeled dataset and added to the labeled dataset, while being removed from the unlabeled dataset. This ensured the utilization of all labeled and unlabeled data for fine-tuning the global model before the final global update.

\subsection{Experimental Results of LP-FL on IMDB and Yelp}
\label{sec:LP_FL_result}

\begin{table}[!ht]

\caption{LP-FL results}
\label{tab:LP_FL_result}
\centering

\renewcommand{\arraystretch}{1.5}
\begin{tabular}{ p{1.5cm}<{\centering} | p{2.5cm}<{\centering}  p{2cm}<{\centering}  p{2cm}<{\centering}  p{2cm}<{\centering}  p{2cm}<{\centering} }

\hline
\multirow{2}*{Dataset} & \multirow{2}*{Labeled Data} & \multirow{2}*{Centralized} & \multicolumn{3}{c}{Our LP-FL (client num)} \\
\cline{4-6}
& & & 2 & 5 & 10 \\
\hline
\multirow{3}*{IMDB} & 1\% & 0.850 & \textbf{0.858} & 0.840 & 0.822 \\
\cline{3-6}
& 5\% & \textbf{0.869} & 0.867 & 0.861 & 0.853 \\
\cline{3-6}
& 10\% & \textbf{0.878} & 0.873 & 0.864 & 0.859 \\
\hline
\hline
\multirow{3}*{Yelp} & 1\% & 0.913 & \textbf{0.922} & 0.903 & 0.900 \\
\cline{3-6}
& 5\% & 0.914 & \textbf{0.927} & 0.919 & 0.911 \\
\cline{3-6}
& 10\% & 0.919 & \textbf{0.929} & 0.925 & 0.919 \\
\hline

\end{tabular}
\vspace{10pt}
\end{table}

\begin{figure}[!ht]
    \centering
    \subfloat[IMDB]{
    \includegraphics[width=0.48\textwidth]{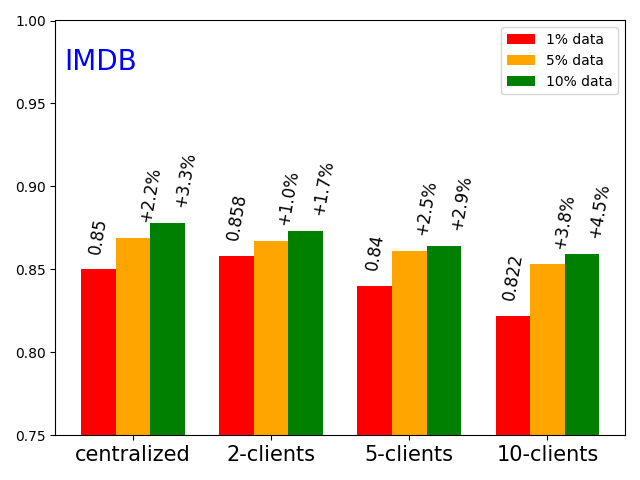}
    \label{fig:imdb_results}
    }
    \hfill
    \subfloat[Yelp]{
    \includegraphics[width=0.48\textwidth]{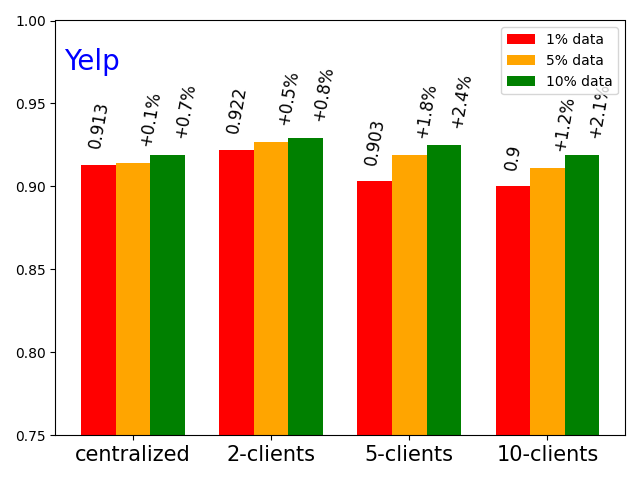}
    \label{fig:yelp_results}
    }
    \caption{LP-FL results on IMDB and Yelp.}
    \label{fig:LP_FL_results}
\end{figure}

\textbf{Result on IMDB and Yelp. } Table~\ref{tab:LP_FL_result} illustrates the outcomes of our method in the context of sentiment analysis task. We have the following observations.

\begin{itemize}[noitemsep,leftmargin=*]
\item \emph{Within the federated environment, our approach consistently achieves a prediction accuracy of over 82\% across the test dataset.} 
\item \emph{A comparative analysis is conducted between our method under the federated environment and the centralized training approach. } Remarkably, our method exhibits minimal decrease in prediction accuracy when there are only two federated participants, demonstrating nearly identical performance to the centralized training approach. 
\item \emph{Moreover, when five and ten client participants engage in the federated training, the prediction accuracy only decrease by approximately 1\% and 2\% respectively.}

\item \emph{In scenarios involving ten federated participants, each contributing a mere 0.1\% of the data (equivalent to 24 data instances), the sample size adequately corresponds to the labeled samples commonly available on mobile devices in practical settings. Nevertheless, our approach consistently yields satisfactory results, with the final model's prediction accuracy only decrease mildly compared to centralized training.}
\end{itemize}

\begin{figure}[!ht]
    \centering
    \subfloat[IMDB]{
    \includegraphics[width=0.48\textwidth]{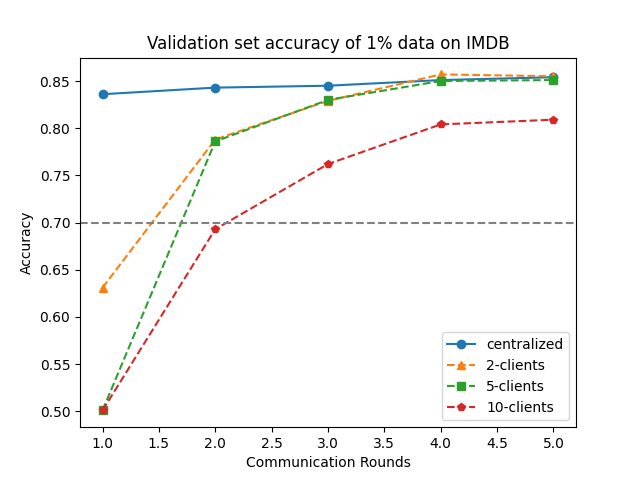}
    \label{fig:imdb_valid_accuracy}
    }
    \hfill
    \subfloat[Yelp]{
    \includegraphics[width=0.48\textwidth]{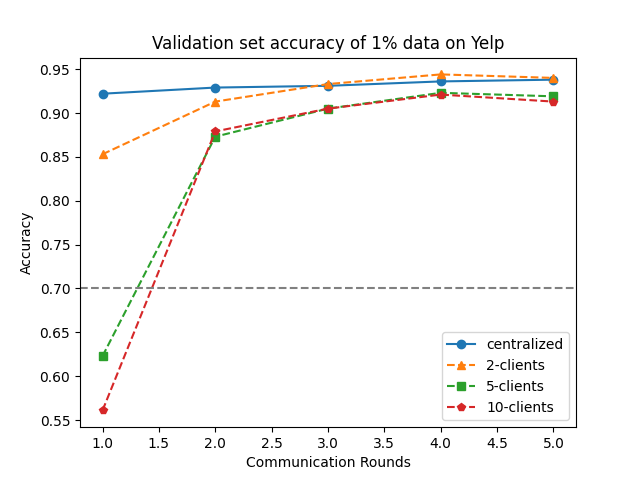}
    \label{fig:yelp__valid_accuracy}
    }
    \caption{Validation set accuracy of 1\% data on IMDB and Yelp.}
    \label{fig:validation_accuracy}
\end{figure}

\textbf{Convergence and Stability of LP-FL. } 
Throughout the experiments, we closely monitor the evolution of the global model's prediction accuracy during each round of global training. 

Notably, we show in Figure ~\ref{fig:validation_accuracy} an extreme case of using a minimal of 1\% subset of the IMDB and Yelp datasets. Although the accuracy was low in the initial 1-2 rounds, it quickly rises to a reasonable level by just a few more rounds of training.


This nice property of quick convergence can be partly attributed to our low-parameter methodology of training. That is, with limited training samples, we seek to fine-tune only about 30\% of the total model parameters, preventing over-fitting to the small number of samples. 

By employing our proposed iterative and semi-supervised methodology, we can enhance the utilization of unlabeled data in the presence of scarce labeled data. We establish a criterion that the prediction accuracy of the global model after FedAvg exceeds 70\% before using it to annotate the unlabeled data. This approach allows us to capitalize on a greater quantity and higher quality of data for fine-tuning purposes. Furthermore, we ensure the completion of annotating all unlabeled data prior to initiating the final global update, thus ensuring consistency in the training data volume and fully exploiting all available local data. 

As shown in Table~\ref{fig:LP_FL_results} , the final prediction accuracy of the global model of LP-FL, when compared to centralized training (third column), exhibits a performance loss of 2.8\% (0.822 vs. 0.850) on IMDB and 1.3\% (0.900 vs. 0.913) on Yelp using 1\% subset. In contrast, if we increase the labeled data to 10\%, the performance loss diminishes to 1.9\% (0.859 vs. 0.878) on IMDB and becomes negligible on Yelp. These trends align with our expectations and falls within an acceptable range, thereby demonstrating the stability of LP-FL.





\begin{table}[ht]
\caption{Comparison with FP-CT and FP-FL}
\label{tab:comparison_with_full}
\centering
\renewcommand{\arraystretch}{1.5}
\begin{tabular}{ p{1.5cm}<{\centering} | p{1.5cm}<{\centering} p{1.5cm}<{\centering}  p{2.5cm}<{\centering} p{1.5cm}<{\centering}  p{2.5cm}<{\centering}}

\hline
\multirow{2}*{Dataset} & \multirow{2}*{Labeled} & \multicolumn{2}{c}{LP-} & \multicolumn{2}{c}{FP-} \\
\cline{3-6}
& & CT & FL(2-clients) &  CT & FL(2-clients) \\
\hline
\multirow{2}*{IMDB} & 1\% & 0.850 & \textbf{0.858} & 0.855 & 0.850 \\
\cline{3-6}
& 5\% & 0.869 & 0.867 & \textbf{0.872} & 0.864 \\
\hline
\hline
\multirow{2}*{Yelp} & 1\% & 0.913 & \textbf{0.922} & 0.914 & 0.916 \\
\cline{3-6}
& 5\% & 0.914 & \textbf{0.926} & 0.922 & 0.924 \\
\hline
\end{tabular}
\end{table}

\begin{figure}[!ht]
    \centering
    \subfloat[IMDB]{
    \includegraphics[width=0.48\textwidth]{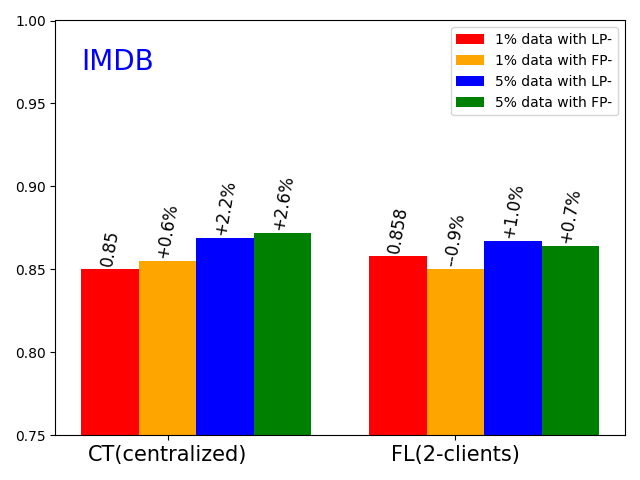}
    \label{fig:imdb}
    }
    \hfill
    \subfloat[Yelp]{
    \includegraphics[width=0.48\textwidth]{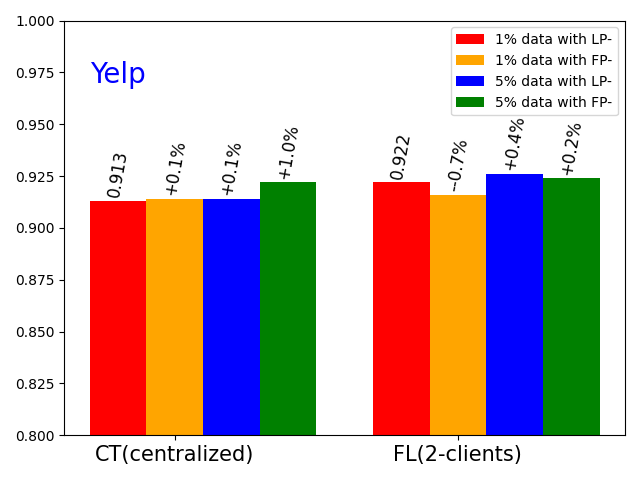}
    \label{fig:yelp}
    }
    \caption{Comparison with FP-CT and FP-FL.}
    \label{fig:comparison_with_full}
\end{figure}

\subsection{Ablation Studies}
We compare the performance of our method with Full-Parameter Centralized Training (FP-CT) and Full-Parameter Federated Learning (FP-FL) in Table~\ref{tab:comparison_with_full}. Based on the comparisons, we draw the following conclusions:

\textbf{FP-CT is better than LP-CT. }As expected, the prediction accuracy of LP-CT is slightly lower than that of FP-CT, although the decrease is not substantial. The smallest observed difference is only 0.1\% (0.913 vs. 0.914), while the larger discrepancy is merely 0.8\% (0.914 vs. 0.922), as shown in Table~\ref{tab:comparison_with_full} column of LP-CT and FP-CT. The reason behind this is simple: FP-CT utilizes the complete set of model parameters for fine-tuning, enabling it to acquire more knowledge compared to LP-CT, which utilizes only approximately 30\% of the parameters. Consequently, FP-CT exhibits superior performance.

\textbf{LP-FL is better than FP-FL. }Unexpectedly, the prediction accuracy of the LP-FL consistently outperforms the FP-FL in the federated environment (0.858 vs. 0.850 and 0.867 vs. 0.864 on IMDB, 0.922 vs. 0.916 and 0.926 vs. 0.924 on Yelp). This observation contradicts the majority of related work in centralized training. We attribute these findings to the semi-supervised nature of our approach. We annotate the unlabeled dataset only after achieving a high level of model prediction accuracy. However, there are still some erroneous samples present. Additionally, the large parameter size of the language model contributes to its complexity, making it more prone to memorizing noise. Under FP-FL, the erroneous annotations introduce more noise to the model, adversely affecting prediction accuracy. On the other hand, LP-FL only fine-tunes around 30\% of the parameters, reducing the model's complexity and minimizing the impact of erroneous annotations. As a result, over-fitting to noise caused by erroneous annotations is reduced under LP-FL, leading to better performance compared to FP-FL.

\textbf{LP-FL is equal or better than LP-CT. }The presence of noise induced by erroneous annotations, derived from the semi-supervised nature, is also observable in LP-CT and LP-FL. It has been noted in our investigations that the prediction accuracy of LP-CT is also lower than that LP-FL in certain settings (0.850 vs. 0.858 on IMDB, 0.913 vs. 0.922 and 0.914 vs. 0.926 on Yelp) as shown in Table~\ref{tab:comparison_with_full} column of LP-CT and LP-FL. Although LP-CT reduces model complexity and mitigates the impact of erroneous annotations, the quantity of erroneous annotations still affects prediction accuracy. Centralized training models have access to a larger unlabeled dataset, leading to more erroneous annotations after labeling. In contrast, federated participants have smaller local unlabeled datasets, resulting in fewer erroneous annotations. Given the same model complexity, a larger number of erroneous annotations increases the likelihood of over-fitting to incorrect labels, resulting in lower prediction accuracy.

\textbf{FP-FL is equal or better than FP-CT. }FP-FL demonstrates prediction accuracy that is comparable to, and sometimes even surpasses, that of FP-CT. This finding provides additional evidence that in a semi-supervised setting, a greater number of erroneous annotations result in increased noise, thereby impacting the model's prediction accuracy. Due to the relatively fewer erroneous annotations encountered by individual models in the federated environment compared to centralized training, FP-FL occasionally outperforms FP-CT.





\section{Conclusion} 
We have validated that pre-trained large language models (LLMs) can attain reasonable classification accuracy in sentiment classification tasks with minimal labeled samples across different clients with federated learning paradigm. 
Our proposed Low-Parameter Federated Learning (LP-FL) permits federated clients to assign soft labels to unlabeled data, utilizing the evolving knowledge from the global model. 
Given the inherent constraints of mobile devices, such as limited computational resources and unreliable network environments, we achieve comparable, and sometimes superior performance to Full-Parameter Federated Learning (FP-FL) by fine-tuning and federating only a fraction of the model parameters at each local client.  Our FP-FL approach establishes an effective learning framework for leveraging LLMs in the FL environment and shows promising applications over mobile devices.

{
\bibliographystyle{splncs04}
\bibliography{egbib,fed,plm}
}

\end{document}